    \documentclass[%
 reprint,
 superscriptaddress,
 amsmath,amssymb,
 aps,
 prx,
floatfix,
longbibliography 
]{revtex4-2}

\usepackage{epsfig}
\usepackage{amssymb}
\usepackage{amsmath}
\usepackage{amsfonts}
\usepackage{braket}
\usepackage[T1]{fontenc}
\usepackage{bm}
\usepackage{graphicx}
\usepackage{xcolor}
\usepackage{verbatim}
\usepackage{soul}
\usepackage{tikz}
\usetikzlibrary{matrix,shapes,arrows,positioning,chains}
\usetikzlibrary{shapes.geometric}
\usetikzlibrary{shapes.arrows}
\usepackage{array}
\usepackage{enumitem}
\usepackage{color}
\usepackage{physics}
\usepackage[export]{adjustbox}
\usepackage[percent]{overpic}
\usepackage[normalem]{ulem}

\usepackage{bbold}

\usepackage{float}
\usepackage{dcolumn}
\usepackage{bm}
\usepackage[pdftex,colorlinks=true]{hyperref}
\hypersetup{
  colorlinks=true,
  linkcolor=blue,
  urlcolor=cyan,
}
\usepackage{tikz}
\usetikzlibrary{decorations.pathreplacing}

\newcommand{\stkout}[1]{\ifmmode\text{\sout{\ensuremath{#1}}}\else\sout{#1}\fi}
\newcommand{\eqn}[1]{\begin{equation} #1 \end{equation}}
\newcommand{\eqa}[1]{\begin{align} #1 \end{align}}
\newcommand{\nn}{\nonumber}

\newcommand{\sectionn}[1]{\textit{#1---}}

\newcommand{\mH}{\mathcal{H}}
\newcommand{\mD}{\mathcal{D}}
\newcommand{\mZ}{\mathcal{Z}}
\newcommand{\mS}{\mathcal{S}}

\newcommand{\pd}{\partial}

\newcommand{\bn}{\boldsymbol{n}}

\newcommand{\bz}{\boldsymbol{\hat{z}}}

\newcommand{\bl}{\boldsymbol{l}}

\newcommand{\m}{\boldsymbol{m}}
\newcommand{\bA}{\boldsymbol{A}}

\newcommand{\bsigma}{\boldsymbol{\sigma}}
\newcommand{\bQ}{\boldsymbol{\Theta}}

\newcommand{\hbm}{\hat{\boldsymbol{m}}}
\newcommand{\hbn}{\hat{\boldsymbol{n}}}
\newcommand{\hbq}{\hat{\boldsymbol{\theta}}}
\newcommand{\hbf}{\hat{\boldsymbol{\phi}}}

\begin{document}

\title{A Generalised Haldane Map from the Matrix Product State Path Integral \\to the Critical Theory of the $J_1$--$J_2$ Chain}

\author{F. Azad}
\thanks{These authors contributed equally to this work}
\affiliation{Electrical Engineering and Computer Science Department, Technische Universit\"{a}t Berlin, 10587 Berlin, Germany}
\affiliation{London Centre for Nanotechnology, University College London, Gordon St., London, WC1H 0AH, United Kingdom}

\author{Adam J. McRoberts}
\thanks{These authors contributed equally to this work}
\affiliation{International Centre for Theoretical Physics, Strada Costiera 11, 34151, Trieste, Italy}
\affiliation{Max Planck Institute for the Physics of Complex Systems, N\"{o}thnitzer Str. 38, 01187 Dresden, Germany}

\author{Chris Hooley}
\affiliation{Max Planck Institute for the Physics of Complex Systems, N\"{o}thnitzer Str. 38, 01187 Dresden, Germany}

\author{A.~G. Green}
\affiliation{London Centre for Nanotechnology, University College London, Gordon St., London, WC1H 0AH, United Kingdom}

\date{\today}
\begin{abstract}
\noindent
We study the $J_1$-$J_2$ spin-$1/2$ chain using a path integral constructed over matrix product states (MPS). By virtue of its non-trivial entanglement structure, the MPS ansatz captures the key phases of the model even at a semi-classical, saddle-point level, and, as a variational state, is in good agreement with the field theory obtained by abelian bosonisation. 
Going beyond the semi-classical level, we show that the MPS ansatz facilitates a physically-motivated derivation of the field theory of the critical phase: by carefully taking the continuum limit---a generalisation of the Haldane map---we recover from the MPS path integral a field theory with the correct topological term and emergent $SO(4)$ symmetry, constructively linking the microscopic states and topological field-theoretic structures. Moreover, the dimerisation transition is particularly clear in the MPS formulation---an explicit dimerisation potential becomes relevant, gapping out the magnetic fluctuations.    
\end{abstract}
\maketitle

The attempt to combine variational methods with quantum field theory has a long history~\cite{Feynman_Variational}, with a significant advance coming from the formulation of matrix product states directly in the continuum limit~\cite{PhysRevLett.104.190405,PhysRevB.88.085118}. An alternative approach, that follows the traditional route of condensed matter physics more directly~\cite{Haldane:1982tu,Affleck:1987mw}, is to develop a spatially discrete path integral over an appropriate variational manifold~\cite{green2016feynman} and then to construct a field theory as a continuum limit. The question of how the topological terms of conventional field-theoretical approaches emerge in this prescription has up to now been unclear. 

The $J_1$-$J_2$ chain has provided an archetypal model in which to investigate the role of topology in quantum systems, and harbours a rich phase diagram. Semi-classically, an incommensurate helimagnetic phase interpolates between the ferromagnetic and antiferromagnetic (N\'eel) phases, and a plethora of numerical and analytical approaches have revealed further features~\cite{furukawa2012ground,majumdar1969next,Majumdar_1969,OKAMOTO1992433,Nomura_1994,white1996dimerization,PhysRevLett.81.910}. 
We will be most concerned with the transition in the quantum $S = \frac{1}{2}$ chain between a critical antiferromagnetic phase and the dimer phase, where competing first- and second-neighbour antiferromagnetic terms drive adjacent spins into singlets.

Under abelian bosonisation~\cite{Haldane:1982tu,Affleck:1987mw,von1998bosonization,giamarchi2003quantum}, this transition appears to be of Kosterlitz-Thouless (KT) character. In obscuring the $SU(2)$ symmetry, however, this misses the spacetime topology of the quantum states, and the fact that topological defects in one phase are bound to the charges of the other---the boundary between two singlet covers has a single (delocalised) spin. Field-theoretically, this is encoded in additional topological Wess-Zumino (WZ) terms~\cite{tanaka2002quantal,tanaka2005many,senthil2006competing} (generalisations of the Haldane $\Theta$-term~\cite{haldane19883}), which measure the winding of the joint N\'eel-singlet order parameter during the tunneling events (instantons) that drive the transition out of the dimer phase.

These ideas are central to the understanding of deconfined quantum criticality~\cite{Senthil:2004qi} and underpin a rich network of dualities between different two-dimensional quantum states. In the $J_1$-$J_2$ model, these terms may be derived via a mapping of the spins to fermionic degrees of freedom~\cite{tanaka2002quantal,tanaka2005many,senthil2006competing}. In this letter, we give an alternative derivation working with the original spin degrees of freedom, mirroring the derivation of the $\Theta$-term in the antiferromagnet~\cite{haldane19883}, and drawing an explicit, constructive connection between tensor network states and field-theoretic topological terms.

A matrix product state (MPS) ansatz encompassing the key physics of the different phases is the basis of our approach---even as a variational wavefunction, it captures a continuous phase transition between the critical and dimer phases. Reparametrising in terms of the order parameters of these phases, we construct the path integral over the MPS ansatz~\cite{green2016feynman}, and demonstrate how the resulting effective action recovers the non-linear sigma model (NLSM) of the critical phase, with the $SO(4)$ WZ term (cf.~\cite{haldane19883}), explicitly linking these topological field-theoretic structures to the microscopic states. 
Further, the MPS field theory clarifies the nature of the dimerisation transition, with an explicit potential term which breaks the $SO(4)$ symmetry and opens the gap.

\begin{figure*}[t]
\includegraphics[width=0.40\textwidth]{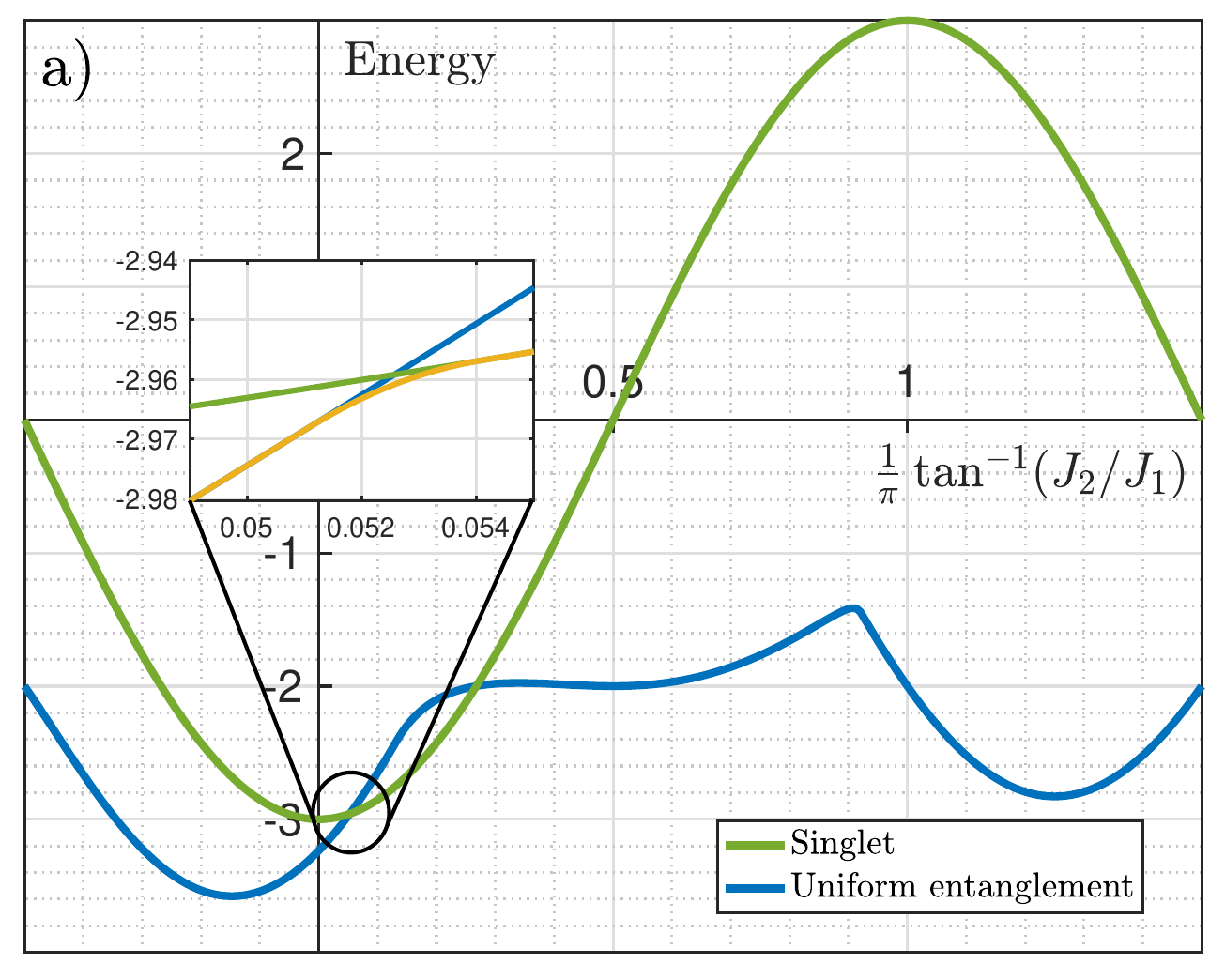}
\includegraphics[width=0.40\textwidth]{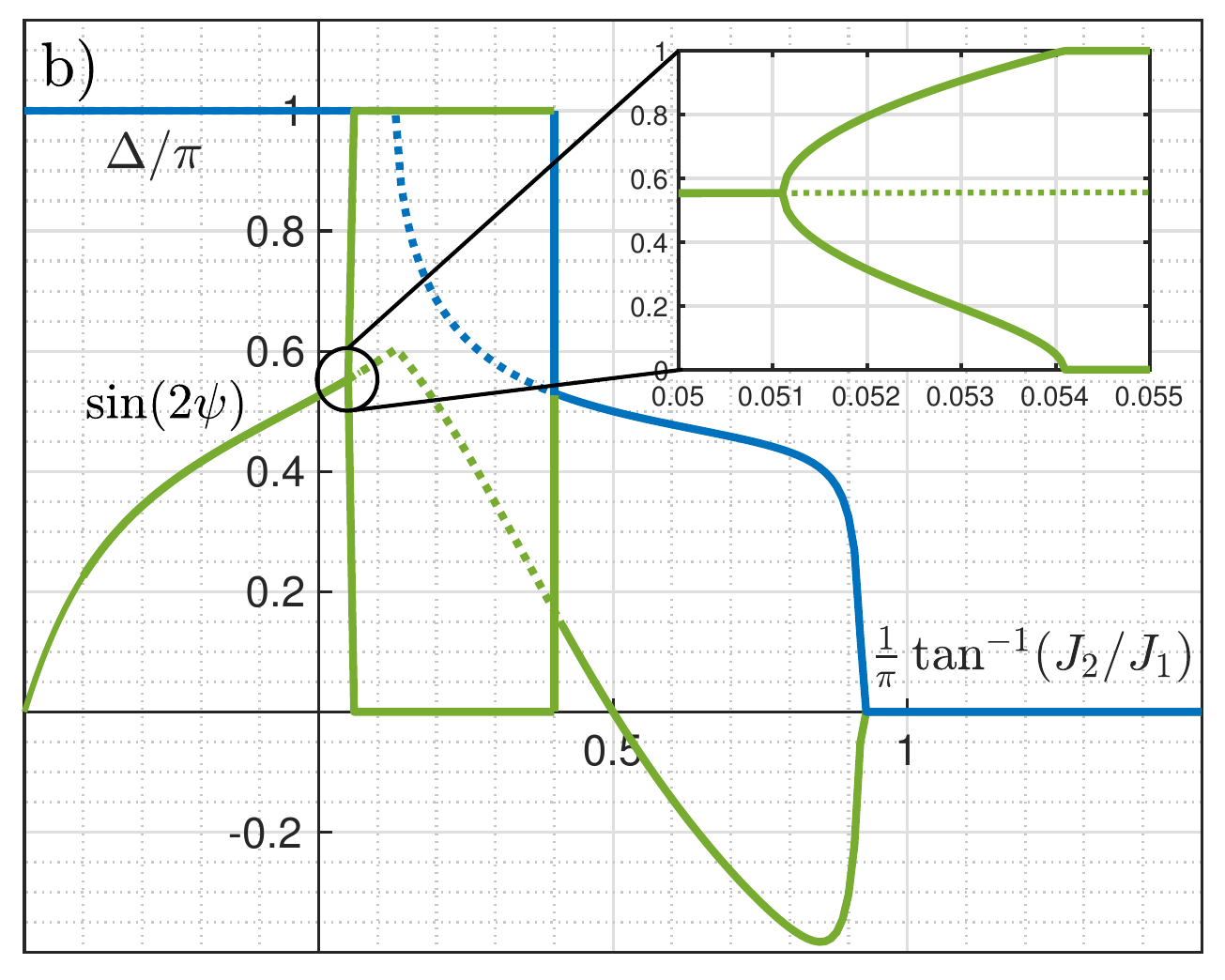}
\caption{{\bf Saddle point phase diagram} as a function of  $\tan^{-1}(J_2/J_1)/\pi$. (a) Variational ground state energy: the blue curve shows the energy of the uniform entanglement saddle point ($\psi_A = \psi_B$), and the green curve the energy of a singlet cover. Near the crossover (inset), another saddle point solution emerges (see \cite{supplemental}), so that the transition is continuous even at this level. (b) The helimagnetic pitch $\Delta/\pi$ (blue) and the entanglement $\sin(2 \psi)$ (green). The entanglement is uniform except in the singlet phase, and near the transition (inset). The singlet phase suppresses the development of an incommensurate spiral (dashed curves show the optimum uniform entanglement state).
}
\label{fig:PhaseDiagram}
\end{figure*}

\sectionn{Model and MPS ansatz}\label{Sec:ModelandAnsatz}
The Hamiltonian of the \mbox{$J_1$-$J_2$} chain consists of competing Heisenberg interactions between nearest and next-nearest neighbours,
\begin{equation}
\hat{H} = 
 \sum_{i} 
 \left( 
 J_1 \hat {\bm \sigma}_i \cdot \hat {\bm \sigma}_{i+1} 
 +
 J_2 \hat {\bm \sigma}_i \cdot \hat {\bm \sigma}_{i+2} 
 \right),
\label{eq:J1J2Model}
\end{equation}
where $\hat{\boldsymbol{\sigma}} = (\hat{\sigma}^x, \hat{\sigma}^y, \hat{\sigma}^z)$ is the vector of Pauli matrices. 
We introduce our variational state as an MPS ansatz with bond-dimension two,
,
\eqa{
\ket{\Psi} &= \sum_{\{\sigma\}} \underset{i}{\scalebox{0.9}{\(\bigotimes\)}} \left( A_{[i]}^{\sigma_i} \ket{\sigma_i\hbn_i} \right), \;
A_{[i]}^{+} = \begin{pmatrix} \cos\psi_i  & 0 \\ 0 & i e^{i\xi_i}\sin\psi_i \end{pmatrix}, \nn \\
&\;\;\;\;\;\;\;\;\;\;\;\;A_{[i]}^{-} = e^{i\pi/4} \begin{pmatrix} 0  & \cos\psi_i \\ i e^{i\xi_i}\sin\psi_i & 0 \end{pmatrix},
\label{eq:Ansatz}
}
where $\ket{\hbn}$ represents a spin-coherent state polarised in the direction of the unit vector $\hbn$. (we discuss the gauge-fixing in the supplementary \cite{supplemental}). The angle $\xi$ allows tuning between singlet and $S=0$ triplet configurations and will generally not appear in the following.

This state captures the key phases of the $J_1$-$J_2$ model. In particular, it can represent a product state antiferromagnet with $\psi = 0$ or $\pi/2$, and $|{\hbn}_i\rangle = |(-1)^i {\hbn} \rangle$, and the singlet covers are obtained by setting  
${\psi_{2i}= \pi/4}$,  ${\psi_{2i-1} = 0}$ (or vice-versa), with ${|{\hbn}_i\rangle = |(-1)^i{\hbn}\rangle}$.

\sectionn{Saddle-point phase diagram}\label{Sec:EnergeticsAndPhase}
Before turning to the connections to the field theory, let us first discuss the phase diagram that results from the ansatz \eqref{eq:Ansatz}. The finite, two-site correlation length implies that $\langle\hat{H}\rangle$ can be calculated and manipulated analytically; Eq.~(\ref{eq:Ansatz}) is in left canonical form with left-environment $\mathbb{1}$ and right-environment ${R_i= ({\mathbb{1}} + \sigma^z \cos 2\psi_{i+1})/2}$.

In order to simplify the optimisation, we restrict $\hat\bn$ to lie in the $xy$-plane with a constant pitch angle $\Delta \in [0, \pi]$, such that $\hat\bn_i \cdot \hat\bn_{i+1} = \cos\Delta$. 
Further, we assume that the entanglement parameters $\psi_i$ will at most alternate between two values $\psi_A$ on even sites and $\psi_B$ on odd sites.
Now, the expectation value of the Hamiltonian (\ref{eq:J1J2Model}), over the MPS ansatz (\ref{eq:Ansatz}) with these restrictions, is
\eqa{
E = &J_1 \Bigl[ (\cos^2  2\psi_A + \cos^2  2 \psi_B)  \cos \Delta \nn \\
&\;\;\;\;\;\;\;\;\;- \sin[ 2(\psi_A+\psi_B)] \left( 1 - \cos \Delta  \right) \Bigr] \nn \\
&+ J_2 \Bigl[2 \cos^2 2 \psi_A \cos^22 \psi_B \cos2 \Delta \nn \\
&\;\;\;\;\;\;\;\;\;\;\;\;+ \sin 4 \psi_A \sin 4 \psi_B (1+ \cos2 \Delta)/4
\Bigr].
\label{eq:EnergyInAnsatz}
}
The variational (saddle-point) phase diagram (Fig.~\ref{fig:PhaseDiagram}) follows by minimising Eq.~(\ref{eq:EnergyInAnsatz}) over $\psi_A$, $\psi_B$ and $\Delta$.
The pitch angle follows the classical result for most of the phase diagram---an (incommensurate) helimagnetic phase with $\Delta = \cos^{-1}[-J_1/(4 J_2)]$ interpolates between ferromagnetic and N\'eel order at $J_1/J_2 < -4$ and $J_1/J_2 > 4$ respectively, dressed with some uniform entanglement. The helimagnetic order is suppressed, however, by the dimerised singlet phase for $0.1619 \lesssim J_2/J_1 \lesssim 1.317$. 

The transition between the N\'eel and singlet phases occurs at $J_2/J_1\approx1/6$. On the scales indicated in Figs.~\ref{fig:PhaseDiagram}(a) \& (b), the singlet phase appears to be formed at an abrupt first order discontinuity in the parameters of the MPS ansatz; however, zooming in on the region around this point reveals two continuous transitions (at saddle-point level): the first at $J_2/J_1\approx0.1619$, where the translation symmetry is first broken; and the second at $J_2/J_1 = 3-2 \sqrt{2} \approx 0.1716$, where the singlet state is fully formed (see the supplementary \cite{supplemental} for the analytic details). 

\begin{figure}[t]
    \centering
    \includegraphics[width=0.8\columnwidth]{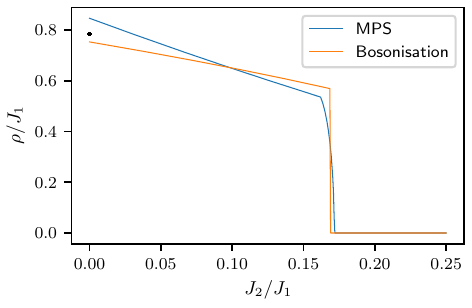}
    \caption{{\bf The spin stiffness} $\rho$ as a function of $J_2/J_1$ across the N\'{e}el-dimer transition, obtained from the saddle point configuration of the MPS ansatz (\ref{eq:Ansatz}) and the bosonised field theory \cite{supplemental}. There is reasonable agreement between the two methods, with both exhibiting the jump at $J_2 = J_2^c$. The black point is the exact (Bethe-ansatz) value, $\pi/4$, at $J_2 = 0$.}
    \label{fig:spin_stiffness}
\end{figure}

This in-plane optimisation of the MPS ansatz invites a comparison with abelian bosonisation, which predicts a KT dimerisation transition around the same point $J_2/J_1 \approx 1/6$~\cite{Haldane:1982tu, Affleck:1987mw, supplemental}. Whilst this picture will be modified by the topological terms, we still expect a universal jump in the spin stiffness $\rho$ (which we derive in the supplementary \cite{supplemental}). The MPS ansatz \eqref{eq:Ansatz}, on the other hand, furnishes us with an estimate of $\rho$ much more straightforwardly: we simply use the dependence of the energy upon the pitch $\Delta$ to evaluate the resistance to inducing a twist in the magnetic order. That is,
\begin{equation}
\rho = \langle N \rangle \left.\frac{\partial^2 E}{\partial \Delta^2}\right|_{\Delta=\pi},
\end{equation}
where $\langle N\rangle  = \cos(2\psi_A)\cos(2\psi_B)$ is the N\'{e}el order parameter. We show the spin stiffness in Fig.~\ref{fig:spin_stiffness}, and find that, \textit{even at saddle point level}, the universal jump associated to the KT transition is visible---though it occurs between the split transitions, rather than discontinuously---and broadly agrees with the field-theoretic estimate \footnote{Optimising over the full bond-order two MPS manifold produces a better variational state, but the point of the MPS ansatz \eqref{eq:Ansatz} is that it can be manipulated analytically---allowing it to connect to the field theory.}.

The rest of our treatment focuses on this dimerisation transition, for which later work~\cite{tanaka2002quantal,tanaka2005many,senthil2006competing} went beyond a purely KT treatment and showed the importance of WZ terms in the field theory---they encode the binding of spins to domain walls between singlet covers. We will show in the following that the MPS ansatz captures this physics directly, in the structure of the spin wavefunction. We note that a higher bond order ansatz would result in better saddle-points for the path integral, but we stress that this is not a variational method -- the spirit of our approach is to select the lowest bond dimension that captures the qualitative features of the competing orders at the saddle-point level, ensuring both that the path integral is analytically tractable and that the correct critical theory can be constructed in simple perturbative corrections to the saddle-points.

\begin{figure}[t]
\includegraphics[width=0.8\columnwidth]{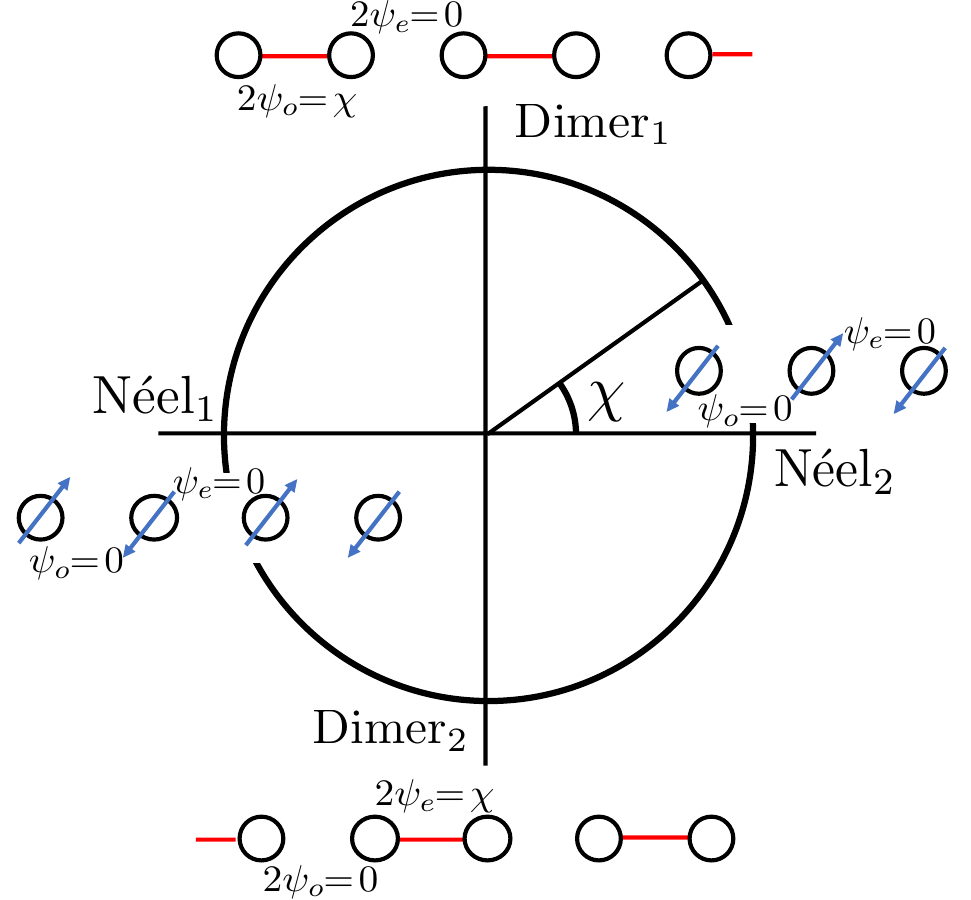}\vspace{0.3cm}
\begin{tikzpicture}
    \foreach \x in {2,3,4,5,6,7,8,9} {
        \fill[black] (\x,0) circle (2pt);
    }
    \draw[blue] (2,0) -- (3,0);
    \draw[black, ->] (3.9, -0.3) -- (4.1, 0.3);
    \draw[red] (5,0) -- (6,0);
    \draw[black, <-] (6.9, -0.3) -- (7.1, 0.3);
    \draw[blue] (8,0) -- (9,0);

    \draw[black, dotted] (1.5,0) -- (1.7,0);
    \draw[black, dotted] (9.2,0) -- (9.5,0);

    \draw (2.5,-0.2) node[below] {$\chi_e = \chi_o$};
    \draw (2.5,+0.2) node[above] {$\chi > 0$};
    \draw (4,-0.2) node[below] {$\chi_e = 0$};
    \draw (5.5,-0.2) node[below] {$\chi_o = \chi_e$};
    \draw (5.5,+0.2) node[above] {$\chi < 0$};
    \draw (7,-0.2) node[below] {$\chi_o = 0$};
    \draw (8.5,+0.2) node[above] {$\chi > 0$};
    \draw (8.5,-0.2) node[below] {$\chi_e = \chi_o$};
\end{tikzpicture}
\caption{{\bf Order Parameter:} The parameters of the MPS ansatz (\ref{eq:Ansatz}) must be patched together in order to construct the $SO(4)$ order parameter. The polar angle $\chi$ (restricted to the range $\chi \in [-\pi/2, \pi/2]$ to avoid a double cover in the usual way) is introduced so that between $-\pi/2$ and $0$ it describes entanglement on even bonds, and between $0$ and $\pi/2$ entanglement on the odd bonds. A domain wall in the singlet order produces a free spin, where $\chi = 0$.
}
\label{fig:OrderParams}
\end{figure}
%

\sectionn{$SO(4)$ order parameter and continuum limit}\label{Sec:ContinuumLimit}
The first step is to construct a local $SO(4)$ order parameter that treats the N\'eel and dimer order on an equal footing.
In terms of the MPS, the N\'eel order parameter is
\begin{eqnarray}
{\bm N} 
&=& (-1)^i\langle \hat {\bm \sigma}_i \rangle 
= \cos (2 \psi_i)  \cos( 2 \psi_{i+1})  \hbn_i.
\label{eq:N}
\end{eqnarray}
And, whilst the singlet states should formally be distinguished by a string order parameter, it suffices here to use a local singlet order parameter
\eqn{
D_i =  (-1)^i \langle \hat \sigma^+_i  \hat \sigma^+_{ i+1} +\hat \sigma^-_i  \hat \sigma^-_{ i+1} - \hat \sigma^+_{i-1}  \hat \sigma^+_{ i} -\hat \sigma^-_{i-1}  \hat \sigma^-_{ i} \rangle
}
valid when ${\hbn}_i \approx -{\hbn}_{i+1}$, where $\hat{\sigma}^+$ and $\hat{\sigma}^-$ are spin raising and lowering operators in the basis $\{\ket{\hbn}, \ket{-\hbn}\}$ (these are the usual raising and lowering operators if $\hbn = \bz$). 
For states given by Eq.~(\ref{eq:Ansatz}), this is, explicitly, 
\begin{eqnarray}
D_i 
&=&  (-1)^i  \left[
 - \sin(2 \psi_{i+1}) \cos (\psi_i-\psi_{i+2}) \cos (\psi_i + \psi_{i+2}) 
 \right.
 \nonumber\\
 & &
 \left.
  + \sin(2 \psi_{i}) \cos (\psi_{i-1}-\psi_{i+1}) \cos (\psi_{i-1} + \psi_{i+1})
  \right].
  \label{eq:LocalD}
\end{eqnarray}
Now, if we allow entanglement on either only even or only odd bonds, we obtain  
\eqa{
&\left[\mathrm{even:}\;\; \psi_{2i+1} = 0\right] && \left[\mathrm{odd:}\;\; \psi_{2i}=0 \right]\nn \\
&\boldsymbol{N}_{2i} = \cos(2\psi_{2i})\hbn_{2i} 
&&\boldsymbol{N}_{2i} = \cos(2\psi_{2i+1})\hbn_{2i} \nn \\
&\boldsymbol{N}_{2i-1} = -\cos(2\psi_{2i})\hbn_{2i-1} 
&&\boldsymbol{N}_{2i+1} = -\cos(2\psi_{2i+1})\hbn_{2i+1}\nn \\
&D_{2i} = \sin(2\psi_{2i}) &&D_{2i} = -\sin(2\psi_{2i+1}) \nn \\
&D_{2i-1} = \sin(2\psi_{2i}) &&D_{2i+1} = -\sin(2\psi_{2i+1}).
}
such that in either case ${\bm N}^2+D^2=1$, and $(D, \boldsymbol{N})$ forms an $SO(4)$ multiplet~\footnote{Note that $\sin(2 \psi)$ and $\cos(2 \psi)$ are swapped compared to the usual polar decomposition of an $SO(4)$ multiplet.}.

We therefore enforce the constraint in the ansatz that no two consecutive $\psi_i$ are nonzero, and introduce a single angle $\chi_i$ which interpolates between the two singlet orders:
\eqn{
\begin{cases}
    2\psi_{2i}= -\chi_{2i}\Theta(-\chi_{2i}) = -\chi_{2i-1}\Theta(-\chi_{2i-1}) \\
    2\psi_{2i+1} = \chi_{2i} \Theta(\chi_{2i}) = \chi_{2i+1} \Theta(\chi_{2i+1}).
\end{cases}}
where $\Theta(x)$ is the Heaviside step function. We show how $\chi$ rotates between the singlet and N\'eel orders in Fig.~\ref{fig:OrderParams}.

We note that this order parameter cannot completely characterise the uniform entanglement saddle-points; those states directly include significant quantum fluctuations which, in the field theory, are encoded as instantons between the two singlet covers.

\sectionn{Effective action}\label{Sec:EffectiveAction}
Now, armed with this joint order parameter, we can construct the field theory as an MPS path integral~\cite{green2016feynman},
\eqn{
\mZ = \int \mD\Psi e^{-\mS[\Psi]},\;\;\; \mS = \int d\tau \bra{\Psi}\ket{\pd_\tau\Psi} + \mH,
}
where ${\mD\Psi=\mD\hbn\mD\chi\mD\xi\cos^2\chi}$ is the functional Haar measure over the restricted MPS ansatz \cite{green2016feynman}. We note that, despite the enforced staggering of $\psi_i$, we are still taking advantage of the fact that the MPS path integral captures entanglement in its saddle-points (the singlet covers). To derive the kinetic and Hamiltonian terms of the action, we can simply set $\psi_{2i+1} = 0$ (the same continuum limit is recovered if we instead set $\psi_{2i} = 0$), though more care is needed for the topological term. We have the Berry phase
\eqa{
\mS_B = \int d\tau \bra{\Psi}\ket{\partial_\tau\Psi} = iS\int d\tau \Bigl[\sum_{i}\dot{\phi}_i\cos\theta_i\cos\chi_{i} \nn \\
+ \sum_{2i}\dot{\xi}_{2i}\left(1-\cos\chi_{2i}\right) \Bigr]
\label{eq:Berry_phase}
}
(where $S = \frac{1}{2}$), and the Hamiltonian,
\eqa{
\mH &= J_1 \sum_{2i} \Bigl[ -w^x_{2i} + (1 + w_{2i}^x)\hbn_{2i-1}\cdot\hbn_{2i} \nn \\ 
&\;\;\;\;\;\;\;\;\;\;\;\;\;\;\;\;\;\; + w^z_{2i} w^z_{2i+2} \hbn_{2i}\cdot\hbn_{2i+1} \;\Bigr] \nn \\
&+ J_2 \sum_{2i} w^z_{2i} w^z_{2i+2}(\hbn_{2i}\cdot\hbn_{2i+2} + \hbn_{2i-1}\cdot\hbn_{2i+1} ),
\label{eq:Hamiltonian}
}
where $w^x=\sin\chi\cos\xi$, $w^z=\cos \chi$ are components of an $O(3)$ vector parametrising the entanglement, and we have set $\psi_{2i+1} = 0$ here and in the $\xi$-term of the Berry phase.

To proceed, we apply Haldane's mapping \cite{auerbach2012interacting} to the spin degrees of freedom,
\eqn{
\hbn_i = (-1)^i \hbm_i \sqrt{1 - \bl_i^2} + \bl_i,
\label{eq:HaldaneMapping}
}
where $\hbm$ is the slowly-varying N\'eel field, and $\bl$ captures the fast fluctuations of the magnetisation, such that,
\eqn{
\hbm_i^2 = 1, \;\;\; \hbm_i\cdot\bl_i = 0, \;\;\; \bl_i^2 \ll 1.
}
We apply this mapping to the Berry phase \eqref{eq:Berry_phase} and Hamiltonian \eqref{eq:Hamiltonian} above, expanding around the saddle points $\chi = \pm\pi/2$ (the singlet covers), and retaining terms up to second order in the fluctuations $\bl$ and $\xi$, and in the derivatives of the slow fields $\hbm$ and $w$. We thus obtain the action \eqref{eq:NLSM}
of the $SO(4)$ NLSM with a topological term and dimerisation potential.

We will give here all of the essential points, and all of the approximations used (see the supplementary \cite{supplemental} for full details). We begin with the topological term: inserting Eq.~\eqref{eq:HaldaneMapping} into Eq.~\eqref{eq:Berry_phase}, we obtain a term
\eqa{
&i\Omega \;=\; iS \int d\tau \sum_{i} (-1)^i\,\dot{\phi}_i \cos\theta_i\cos\chi_{i} \nn \\
\sim iS\int &d\tau dx \left(\frac{\mathrm{sgn}\chi\cos\chi}{2} - \Theta(\chi)\right) \,\epsilon^{\mu\nu}\pd_\mu\cos\theta\,\pd_\nu\phi
\label{eq:BerryPhaseI}
}
(where $\theta$ and $\phi$ are now the angular co-ordinates of the N\'eel field $\hbm$). This is equal to the Wess-Zumino term identified in Ref.~\cite{senthil2006competing} (see also the supplementary \cite{supplemental}),
\eqn{
\Omega_{\mathrm{WZ}}
=
\frac{4 \pi S k}{2 \pi^2}
\int_0^1 d \zeta \int d\tau dx
\,\epsilon_{abcd} 
u^a \partial_\zeta u^b \partial_\tau u^c \partial_x u^d,
\label{eq:BerryPhaseII}
}
where $k\in\mathbb{Z}$ is odd, the factor of $2 \pi^2$ in the denominator is the area of $S^3$, and $u(\zeta, \tau, x)$ is an arbitrary extension of the field $u(\tau,x) = (\sin \chi, \cos \chi {\bm m}) \equiv (u^0, \bm u)$ satisfying 
\begin{eqnarray}
u(\zeta=0,\tau,x) &=& (1,0,0,0) \nonumber\\
u(\zeta=1,\tau,x) &=& u(\tau,x).
\end{eqnarray}
The extension allows a co-ordinate free expression of the WZ term, but $\Omega$ and $\Omega_{\mathrm{WZ}}$ produce the same functional weight for any configuration $u(\tau, x)$ in the path integral.

In the MPS treatment, the internal structure of the ansatz constrains domain walls between different singlet covers to contribute to the overall Berry phase as a free spin; the topological term imposes this same constraint in the field theory.

The kinetic terms follow by integrating out the fast fields, $\xi$ and $\bl$. Integrating out $\xi$ is straightforward after expanding Eq.~\eqref{eq:Hamiltonian} to quadratic order. To integrate out $\bl$ we neglect any terms $\sim \cos^2\chi \, \bl^2$, but otherwise it proceeds as in the standard Haldane mapping \cite{auerbach2012interacting} (see also the supplementary \cite{supplemental}). We obtain kinetic terms $(\pd_\tau w^z)^2$ and $(w^z)^2 (\pd_\tau \hbm)^2$, respectively, with the same prefactor $1/(32J_1)$.

To deal with the Hamiltonian, we simply expand in the gradients of $\hbm$ and $w$. We note, however, that fluctuations of $\hbm$ are heavily suppressed on the entangled bonds---this means that gradients of $\hbm$, like the gradients of $w^z$, occur over two lattice spacings, not one. Accounting for this fact ensures that both terms, $(\pd_x w^z)^2$ and $(w^z)^2(\pd_x\hbm)^2$, appear with the same prefactor, $J_1 - 2J_2$.

Finally, we switch to the standard angular co-ordinates on $S^3$, i.e., we define $\alpha = \pi/2 - \chi$, in terms of which $(w^z)^2 = \sin^2\alpha$ and $(\pd w^z)^2 \sim (\pd\alpha)^2$. We have, then,
\eqa{
\mS &= i\Omega + \int d\tau dx \biggl[ \frac{1}{32J_1}\left((\pd_\tau \alpha)^2 + \sin^2\alpha (\pd_\tau\hbm)^2 \right) \nn \\
&+ (J_1 - 2J_2) \left((\pd_x\alpha)^2 + \sin^2\alpha (\pd_x\hbm)^2 \right) + V(\alpha) \biggr],
\label{eq:NLSM}
}
where the dimerisation potential is
\eqn{
V(\alpha) = -\frac{3J_1}{2} + \frac{J_2}{2} - \frac{J_2}{2}\cos2\alpha.
}

This form of the action, derived directly from an MPS parametrisation of the spin states \eqref{eq:Ansatz}, makes the physics of the transition particularly transparent: if $J_2 < J_2^c$, $V(\alpha)$ is irrelevant, the $SO(4)$ symmetry emerges in the infrared, and the topological term ensures the theory remains gapless; if $J_2 > J_2^c$, $V(\alpha)$ is relevant, and the $SO(4)$ symmetry is broken in favour of dimer order.

\sectionn{Discussion and outlook}\label{Sec:Discussion}
In this letter, we have introduced an MPS ansatz that captures the key physics of the $J_1$-$J_2$ chain, encompassing ferromagnetic, N\'eel, spiral, and dimer orders. Even at saddle-point level, it reproduces the essential features of the deconfined N\'eel-singlet transition at $J_2/J_1\approx1/6$, including the universal jump in the spin stiffness (Fig.~\ref{fig:spin_stiffness})~\footnote{Indeed, if one restricts to helimagnetic phases with uniform entanglement and the pure singlet covers, then the transition is first order at exactly $J_2/J_1=1/6$}.

From this ansatz, we have directly identified a joint N\'eel-dimer order parameter, and constructed the field theory from an MPS path integral~\cite{green2016feynman}. Whilst this is, mechanically, somewhat similar to the construction of the field theory of the Heisenberg antiferromagnet (${J_2 = 0}$) as a coherent state path integral, the use of MPS-valued fields allows us to recover the correct topological term and NLSM~\cite{tanaka2002quantal,tanaka2005many,senthil2006competing}.
Moreover, the nature of the dimerisation transition is remarkably clear in the MPS field theory, where the potential term of an explicit entanglement field flows either to weak- or strong-coupling.

The MPS resums instantons of the coherent state theory and encodes their topological structure locally in the MPS fields. Whilst this connection has been noted previously~\cite{green2016feynman,crowley2014quantum},here we have provided a constructive link: we have shown how the MPS faithfully encodes these topological features, and how this leads directly to the corresponding topological terms in the field theory. 
It is intriguing to speculate that this connection might be used more generally as a method to resum instantons in favour of a higher bond-dimension MPS path integral---the topological term, after all, arises independently of the Hamiltonian. 

The MPS path integral is finding an increasing range of application. It has been used to derive a new treatment of open quantum systems \cite{azad2023},
provides a consistent framework unifying the computation of fluctuation corrections around MPS saddle points \cite{PhysRevB.88.075122,ExcitationAnsatz,VanDamme:2021xqk,LeonticaGreenZeropoint}, has revealed new connections between classical and quantum chaos \cite{hallam2019lyapunov,Leontica:2024lhl}, and with the present work demonstrates the link between tensor-network and field-theoretical treatments of topological structures in quantum wavefunctions. 
Indeed, the effective field theory derived here suggests that the $J_1$-$J_2$ chain has a hitherto unsuspected phase transition \cite{McRHGTopologicalNonlinearSigma} between two distinct critical phases:\ the $O(4)$ non-linear sigma model with topological term suggested by Affleck and Haldane \cite{Affleck:1987mw}, and the recent alternative proposal of Zirnbauer \cite{zirnbauer2024infrared}. 

We anticipate that, with a suitable modification of the underlying MPS ansatz to capture the competing orders in a given system, the MPS path integral will provide a generic method to construct field-theoretic descriptions of quantum phase transitions involving a change in the entanglement structure, such as in higher-spin chains~\cite{xian1993spontaneous,chepiga2016dimerization,yamaguchi2020variety,reja2024topological} or ladder structures~\cite{shelton1996antiferromagnetic,liu2008existence,lamas2015dimerized}. Moreover, whilst generic projected entangled pair states (PEPS) are not efficiently contractible~\cite{schuch2007computational,schuch2008computational}, it may be possible to use sequential circuit ans\"{a}tze~\cite{banuls2008sequentially} to extend the method to two dimensions and capture deconfined quantum criticality~\cite{Senthil:2004qi}.

This work was in part supported by the Deutsche Forschungsgemeinschaft under grants SFB 1143 (project-id 247310070) and the cluster of excellence ct.qmat (EXC 2147, project-id 390858490), by the EPSRC under EP/S021582/1 and EP/I031014, and by the ERU under `Perspectives of a Quantum Digital Transformation'. We acknowledge fruitful discussions in workshops funded by EP/W026872/1.


\bibliography{refs}

\onecolumngrid

  \cleardoublepage
  \begin{center}
    \textbf{\large Supplementary material}
  \end{center}
\setcounter{equation}{0}
\setcounter{figure}{0}
\setcounter{table}{0}
\makeatletter
\renewcommand{\theequation}{S\arabic{equation}}
\renewcommand{\thefigure}{S\arabic{figure}}
\renewcommand{\thetable}{S\arabic{table}}
\setcounter{section}{0}
\renewcommand{\thesection}{S-\Roman{section}}

\setcounter{secnumdepth}{2}

\section*{Contents}

In this Supplementary Material, we give a more detailed analysis of the saddle points of the MPS ansatz; derive an estimate for the spin stiffness from abelian bosonisation, which we compare to the estimate from the MPS ansatz; discuss the dynamical gauge-fixing of the spin coherent states, and show that this does not contribute to the Berry phase; and give a more detailed account of the derivation of the non-linear sigma model from the MPS path integral, including {an explicit demonstration that the correct continuum limit of the lattice Berry phase produces the Wess-Zumino topological term.





\section{Saddle-point analysis of the dimerisation transition}
\label{app:AnalyticalTransition}
 There are three saddle-point equations corresponding to the two-site energy density given in Eq.~(3). Two corresponding to derivatives with respect to the entanglement parameters are given by

\begin{eqnarray}
\frac{\partial E}{\partial \psi_A} 
&=& 
2 J_1(\cos\Delta - 1) \cos(2(\psi_A + \psi_B)) 
- 
 2 (J_1 \cos\Delta + 
    2 J_2 \cos2\Delta \cos^2 2\psi_B) \sin4\psi_A 
 + 
 2 J_2 \cos^2\Delta \cos4\psi_A \sin4\psi_B = 0, 
 \nonumber \\
 \frac{\partial E}{\partial \psi_B} 
 &=& 2 J_1(\cos\Delta - 1) \cos(2(\psi_A + \psi_B)) 
 - 
 2 (J_1 \cos\Delta + 
    2 J_2 \cos2\Delta \cos^2 2\psi_A) \sin4\psi_B
 + 
 2 J_2 \cos^2\Delta \cos4\psi_B \sin4\psi_A = 0,
 \nonumber\\
 \label{eq:saddle_point_eqs_entanglement}
\end{eqnarray}
and one from the derivative with respect to the pitch angle is given by
\begin{eqnarray}
\frac{\partial E}{\partial \Delta} 
&=&
 -\frac{J_1}{2} \sin\Delta \left[2 + \cos4\psi_A + \cos4\psi_B + 
    2\sin2(\psi_A + \psi_B)\right] 
 \nonumber\\
 & &
 -J_2\sin2 \Delta \cos2\psi_A \cos2\psi_B  \left[3\cos2(\psi_A - \psi_B) +  \cos2(\psi_A + \psi_B)\right] = 0.
\label{eq:saddle_point_eq_pitch}
\end{eqnarray}
It is clear that either a ferromagnetic $\Delta = 0$ or an antiferromagnetic $\Delta = \pi$ will solve Eq.~(\ref{eq:saddle_point_eq_pitch}), regardless of the values of $\psi_A$ and $\psi_B$ (solutions involving an incommensurate pitch angle are more involved, and will not be discussed here). Let us focus upon the antiferromagnetic regime, where the N\'{e}el-dimer transition occurs.

With this value for $\Delta$, the saddle-point equations for the entanglement parameters reduce to 
\begin{eqnarray}
\frac{\partial E}{\partial \psi_A} &=& -4 J_1 \cos(2(\psi_A + \psi_B)) + 
 2 (J_1 - 
    2 J_2 \cos^2 2\psi_B) \sin4\psi_A + 
 2 J_2 \cos4\psi_A \sin4\psi_B = 0, 
 \nonumber \\
 \frac{\partial E}{\partial \psi_B} &=& -4 J_1 \cos(2(\psi_A + \psi_B)) + 
 2 (J_1 -
    2 J_2 \cos^2 2\psi_A) \sin4\psi_B + 
 2 J_2 \cos4\psi_B \sin4\psi_A = 0.
 \label{eq:saddle_point_eqs_entanglement_AFM}
\end{eqnarray}

We seek solutions to these reduced saddle-point equations (\ref{eq:saddle_point_eqs_entanglement_AFM}). We first note that the two singlet covers, ${(\psi_A, \psi_B) = (\pi/4, 0)}$ and ${(\psi_A, \psi_B) = (0, \pi/4)}$, are degenerate solutions with $E_S = -3J_1$. 
Deep in the N\'{e}el phase, we expect that the entanglement structure in the ground state will not break any lattice symmetries. Setting $\psi_A = \psi_B = \psi$, both equations reduce, after some simplification, to
\begin{equation}
\frac{\partial E}{\partial \psi} = -4 J_1 \cos4\psi + 
 2 (J_1 - J_2) \sin4\psi = 0.
 \label{eq:uniform_saddle_point_eq}
\end{equation}
The uniform entanglement solution is, therefore,
\begin{equation}
\psi = \frac{1}{4}\arctan\left( \frac{2J_1}{J_1 - J_2} \right),
\label{eq:uniform_saddle_point_sol}
\end{equation}
with energy
\begin{equation}
E_N = -(J_1 - J_2)\left(1 + \sqrt{1 + \left(\frac{2J_1}{J_1 - J_2}\right)^2} \right).
\label{eq:uniform_sol_energy}
\end{equation}
As a check of the quality of this variational ansatz, we can compare this state's energy at ${J_2 = 0}$, ${E_N = -(1 + \sqrt{5})J_1 \approx -3.23601}$, to the exact (Bethe ansatz) ground state at this point, ${-2J_1(1 - 4\ln2) \approx -3.54518}$.

Now, the uniform entanglement state's energy crosses that of the singlet state, $E_S = E_N$, at precisely $J_2/J_1 = 1/6$. This is not, however, where the MPS ansatz (3) predicts the transition to occur. Rather, there is another saddle-point solution to Eqs.~(\ref{eq:saddle_point_eqs_entanglement_AFM}) which interpolates between the uniform and singlet solutions: the transition splits in twain, the entanglement parameters evolve continuously (though not differentiably) as a function of $J_2/J_1$, and the energy is continuously differentiable throughout.

There are two degenerate such solutions, obtained via \textsc{mathematica}, interpolating to either of the singlet covers. The first is
\begin{eqnarray}
\psi_A &=& \frac{1}{2} \arcsin\left(
  \frac{1}{2} \left(\frac{
   1 -6J_2 - 4J_2^2 - 10J_2^3 + 3J_2^4 + 
       4 J_2\sqrt{3J_2^4 - 4J_2^3 + 14J_2^2 + 4J_2 - 1}}{(J_2 - 1)^3 J_2}\right)^{1/2}\right), 
       \nonumber \\
\psi_B &=& -\frac{1}{2} \arctan\left(
  \frac{3J_2 - 1}{1 + J_2} \left(\frac{
   1 -2J_2 - 16 J_2^2 + 2 J_2^3 - J_2^4 + 
       4 J_2\sqrt{3J_2^4 - 4J_2^3 + 14J_2^2 + 4J_2 - 1}}{-1 + 6J_2 + 4 J_2^2 + 10 J_2^3 - 3 J_2^4 - 
       4 J_2\sqrt{3J_2^4 - 4J_2^3 + 14J_2^2 + 4J_2 - 1}}\right)^{1/2}\right),
\end{eqnarray}
where we have set $J_1 = 1$, and the second interchanges $\psi_A$ and $\psi_B$. These transition solutions are only valid ($\psi_A, \psi_B \in \mathbb{R}$) between the two transition points, $J_2^{(-)}$ and $J_2^{(+)}$, where their energy crosses $E_N$ and $E_S$, respectively. Over this region, however, the transition solutions are the lowest energy saddle-point states.

The upper transition point, explicitly, is $J_2^{(+)} = 3 - 2\sqrt{2} > 1/6$. The analytic expression for the lower transition point, whilst it can be expressed using radicals, is much lengthier -- the most concise way of stating it is that $J_2^{(-)}$ is the (unique) positive real root of ${3x^4 - 4x^3 + 14x^2 + 4x - 1}$, which gives $J_2^{(-)} \approx 0.1619 < 1/6$.

\section{Abelian bosonisation and spin stiffness}
\label{app:Bosonisation}
An alternative estimate of the spin stiffness may be obtained from a bosonised field theory. We begin with the linearised $J_1-J_2$ Hamiltonian from Ref.~\cite{Haldane:1982tu},

\eqn{
\hat{H} = \int \frac{dx}{2\pi} \,:\,i J_1\sum_{\eta} \eta\,\hat{\psi}^{\dagger}_{\eta} \pd_x \hat{\psi}_{\eta} 
+ \frac{2J_1}{2\pi}\sum_{\eta\eta'} \hat{\rho}_{\eta} \hat{\rho}_{\eta'}
- \frac{4J_2}{2\pi}\sum_{\eta\eta'} \eta\eta' \hat{\rho}_{\eta} \hat{\rho}_{\eta'}
+ \frac{J_1 - 6J_2}{2\pi} \sum_{\eta} \left(\hat{\psi}^{\dagger}_{\eta}\pd_x \hat{\psi}^{\dagger}_{\eta}\right)\left(\hat{\psi}_{-\eta}\pd_x \hat{\psi}_{-\eta}\right) \,:\,,
}
where $\hat{\psi}_R$, $\hat{\psi}_L$ are chiral fermion fields, the indices $\eta = (+, -)$ correspond to $(R, L)$, the dots $:...:$ denote normal ordering (necessary because the linearisation introduces a Dirac sea of negative energy fermion states), and $\hat{\rho}_{\eta}$ is the density of $\eta$-fermions. The final term is the umklapp term which induces the quantum phase transition in this formulation. Note that some of the coefficients have extra factors of $2\pi$ compared to Ref.~\cite{Haldane:1982tu}, because we are following the conventions of Ref.~\cite{von1998bosonization}.

To these chiral fermions we associate chiral boson fields $\hat{\phi}_{\eta}$, in terms of which the fermion densities are given by $\hat{\rho}_{\eta} = \pd_x \hat{\phi}_{\eta}$, and the fermion fields by vertex operators $\hat{\psi}_{\eta} \sim e^{i\eta \hat{\phi}_{\eta}}$. These chiral bosons have the algebra
\eqn{
\left[\hat{\phi}_{\eta}(x), \pd_x\hat{\phi}_{\eta'}(x')\right] = 2\pi i \eta \delta_{\eta\eta'} \delta(x - x').
}

Defining the total density and current fields,
\eqn{
\hat{\phi} = \frac{\hat{\phi}_R + \hat{\phi}_L}{2},\;\;\;
\hat{\theta} = \frac{\hat{\phi}_R - \hat{\phi}_L}{2},
}
such that $\hat{\phi}$ is canonically conjugate to $\frac{1}{\pi}\pd_x\hat{\theta}(x')$, i.e., 
\eqn{
\left[\hat{\phi}(x), \frac{1}{\pi}\pd_x\hat{\theta}(x')\right] = i\delta(x - x'),
}
the Hamiltonian becomes
\eqn{
\hat{H} = \int \frac{dx}{2\pi} \,:\,\left(J_1 + \frac{4J_1}{\pi}\right)(\pd_x \hat{\phi})^2 + \left(J_1 - \frac{8J_2}{\pi}\right)(\pd_x \hat{\theta})^2 + \frac{J_1 - 6J_2}{2\pi a^2}\cos4\hat{\phi} \,:\,,
\label{eq:H_boson}
}
where we have explicitly reinstated the lattice spacing in the umklapp term. Now, the bare value of the fermion charge stiffness is just the coefficient of the current fluctuations $(\pd_x \hat{\theta})^2$. However, unlike the saddle-point of the MPS ansatz, the bosonised field theory includes ultraviolet contributions from arbitrarily high momentum; we thus identify the physical value of the spin stiffness with the renormalised (infrared) charge stiffness.

We introduce the usual Luttinger liquid parameters $u$ and $K$, where $u$ is the analogue of the Fermi velocity and $K$ is dimensionless. In the Hamiltonian (\ref{eq:H_boson}), $uK$ is the coefficient of $(\pd_x \hat{\theta})^2$, and $u/K$ is the coefficient of $(\pd_x \hat{\phi})^2$. 

To obtain the flow equations, consider the (imaginary time) partition function, 
\eqn{
\mZ = \int \mathcal{D}{\phi}\mathcal{D}{\theta}\, e^{- S[\phi, \theta]},\;\;\; \mS = \int d\tau dx \left[-\frac{i}{\pi}(\partial_{\tau}\phi)(\partial_x\theta) + \mathcal{H} \right],
}
and integrate out the conjugate field $\theta$, leaving the effective sine-Gordon action
\eqn{
\mS = \int dx d\tau \frac{1}{2\pi K} \left( \frac{1}{u} (\partial_{\tau}\phi)^2 + u (\partial_x \phi)^2 \right) + \frac{g}{2\pi a^2} \cos4\phi.
\label{eq:SG_action}
}
The Wilsonian renormalisation procedure of dividing the field into fast modes $\phi_{>}$ and slow modes $\phi_{<}$, and successively integrating out the $\phi_{>}$, may now be performed. Following Ref.~\cite{giamarchi2003quantum}, we have the flow equations
\eqn{
\frac{dK}{dl} = -A K^2 g^2, \;\;\; \frac{dg}{dl} = (2 - 4K)g,
}
for some constant $A > 0$. We can read off the critical value $K_c = 1/2$, below which $g$ always flows to strong coupling. The Luttinger velocity $u$ does not flow under renormalisation -- the action (\ref{eq:SG_action}) is Lorentz covariant, and $u$ is its light speed. 

In principle, we should now compute the infrared values of the couplings. In fact, this is not necessary -- for \textit{all} $J_2 < J_2^c$, the isotropic $J_1$-$J_2$ model lies at a transition between the easy-plane spin-fluid and easy-axis N\'eel state (see Fig.~2 of  Ref.~\cite{Haldane:1982tu}). It follows, then, that the bare values of $K$ and $g$ must lie on the separatrix between the strong and weak-coupling phases, and so $K(l \to \infty) \to K_c = 1/2$, $g(l\to\infty)\to 0$. The transition to the \textit{dimer} state is marked by the bare value of $K$ falling below $K_c = 1/2$, which happens at $J_2 > J_2^c = \frac{3\pi - 4}{32} \approx 1/6$, whereupon $K(l\to\infty)\to 0$; this universal discontinuity in the infrared behaviour of $K$ is the quantum analogue of the universal jump in the spin stiffness in classical Kosterlitz-Thouless transitions. 

With these considerations, the bosonisation estimate for the spin stiffness of the $J_1$-$J_2$ model is
\eqn{
\rho = u K(l\to\infty) = \begin{cases} \frac{1}{2} \sqrt{\left(J_1 + \frac{4J_1}{\pi}\right)\left(J_1 - \frac{8J_2}{\pi}\right)} & J_2 < J_2^c \approx J_1/6, \\ 0 & J_2 > J_2^c \approx J_1/6. \end{cases}
}
We show a comparison of the spin stiffness estimates obtained from the MPS ansatz and bosonisation in Fig.~2 of the main text. The two methods are in reasonable agreement with each other, though neither is exact. Moreover, the MPS estimate is considerably easier to obtain, both conceptually and computationally.

\section{Gauge Fixing}
\label{app:GaugeFixing}
The spin coherent states used in the parametrisation of the MPS ansatz (2) have apparent $U(1)$ gauge  freedom corresponding to rotations of the tangent space basis of the point on $S^2$ specified by $\hbn$. Making this gauge freedom explicit, we have
\eqa{
&\ket{+\hbn} = e^{i\mu/2}\left( \cos\frac{\theta}{2}\ket{\uparrow} + \sin\frac{\theta}{2}e^{i\phi}\ket{\downarrow} \right) \nn \\
&\ket{-\hbn} = e^{-i\mu/2}\left( \sin\frac{\theta}{2}\ket{\uparrow} - \cos\frac{\theta}{2}e^{i\phi}\ket{\downarrow} \right),
\label{eq:CoherentStates}}
where the unit vector parametrising the states is
$\hbn = (\sin\theta\cos\phi, \sin\theta\sin\phi, \cos\theta)$.
We also define
\eqn{
\hbq = \frac{\pd_\theta\hbn}{|\pd_\theta\hbn|} = (\cos\theta\cos\phi, \cos\theta\sin\phi, -\sin\theta), \;\;\; \hbf = \frac{\pd_\phi\hbn}{|\pd_\phi\hbn|} = (-\sin\phi, \cos\phi, 0),
\label{eq:tangent_basis_sup}
}
which provide a basis for the tangent space such that $\{\hbn, \hbq, \hbf\}$ forms a right-handed set. We denote the complex combinations of the tangent space vectors by $\bQ = \hbq + i\hbf$ and $\bQ^* = \hbq - i\hbf$. These combinations show up in the matrix elements of the Pauli matrices with respect to the coherent states,
\eqa{
\bra{\hbn}\bsigma\ket{\hbn} &= \hbn \nn \\
\bra{\hbn}\bsigma\ket{-\hbn} &= -e^{i\mu}\bQ^* \nn \\
\bra{-\hbn}\bsigma\ket{\hbn} &= -e^{-i\mu}\bQ \nn \\
\bra{-\hbn}\bsigma\ket{-\hbn} &= -\hbn,
}
where $\bsigma = (\sigma^x, \sigma^y, \sigma^z)$ is the vector of Pauli matrices. 
Identifying $\bQ = \bQ'= \hbq'+i \hbf'=  -e^{i\mu}\bQ^*$, we see that $\mu$ in Eq.~(\ref{eq:CoherentStates}) is the angle of rotation between $\hbq$, $\hbf$ and $\hbq'$, $\hbf'$.

This redefinition of the tangent space at $\hbn$ allows us to rewrite the expectation value of the Hamiltonian so that the explicit dependence is only upon $\hbn$ and not upon $\bQ$. Taking the expectation value of the Hamiltonian that results from setting $\psi_{2i+1} = 0$ (as in the main text), we have
\eqa{
\mH &= J_1 \sum_{2i} \Bigl[\hbn_{2i-1}\cdot\hbn_{2i} + \cos\chi_{2i} \cos\chi_{2i+2} \hbn_{2i}\cdot\hbn_{2i+1} \nn \\
&\;\;\;\;\;\;\;\;\;\;\;+ \frac{1}{2}\sin\chi_{2i}\left(e^{-i\xi_{2i} - i\mu_{2i-1} - i\mu_{2i}}\bQ_{2i-1}\cdot\bQ_{2i} + e^{i\xi_{2i} + i\mu_{2i-1} + i\mu_{2i}}\bQ^*_{2i-1}\cdot\bQ^*_{2i} \right) \Bigr] \nn \\
&+ J_2 \sum_{2i} \cos\chi_{2i} \cos\chi_{2i+2}(\hbn_{2i}\cdot\hbn_{2i+2} + \hbn_{2i-1}\cdot\hbn_{2i+1}).
\label{eq:AppEnergy}
}
We now perform a series of manipulations that remove the explicit appearance of $\bQ$ and $\bQ^*$ from this expression.
First we note that the appearance of $\mu_{2i-1}-\mu_{2i}$ in Eq.~(\ref{eq:AppEnergy}) reveals that it is in fact not a local gauge invariance when taking account of entangled configurations. We may absorb $\mu_{2i}$ and $\mu_{2i-1}$ into $ \xi_{2i}$, and note that it parametrises the relative phase between $\ket{\uparrow \downarrow}$ and $\ket{\downarrow \uparrow}$ across a bond---and so determines the degree of singlet or triplet correlation. 

Next, we fix $\xi$ to its optimum value. This  can be found by maximising the singlet content across each bond. Geometrically, this optimisation corresponds to  a rotation such that $\hbq'_{2i-1} = \hbq'_{2i} \propto \hbn_{2i-1} \times \hbn_{2i}$, in which case $\hbf_{2i-1}\cdot\hbf_{2i} = \hbn_{2i-1}\cdot\hbn_{2i}$, and we obtain the form (11) for the energy given in the main text. A similar construction is clearly possible if we instead began with $\psi_{2i} = 0$. 

It is instructive to view this simplification of Eq.~(\ref{eq:AppEnergy}) from a complementary perspective.  The change of tangent space basis is equivalent to identifying a rotated polar axis $\hat {\bm z}'_{2i}=\hat {\bm z}'_{2i-1} = \hbq_{2i-1} = \hbq'_{2i}  \propto \hbn_{2i-1} \times \hbn_{2i}$, and the coherent states have a simple expression in terms of spin-up and down states ($| \uparrow_{\hat {\bm z}'} \rangle$ and $| \downarrow_{\hat {\bm z}'} \rangle$, respectively) relative to this polar axis:
$| \hbn_{2i} \rangle = (e^{-i \phi'_{2i}/2} | \uparrow_{\hat {\bm z}'} \rangle + e^{- \phi'_{2i}/2} | \downarrow_{\hat {\bm z}'} \rangle)/\sqrt{2}$. In this case, the expectation values of the spin operators that comprise the middle line in Eq.~(\ref{eq:AppEnergy}) simplify to 
$\langle \hbn | { \bm \sigma} |-\hbn\rangle \cdot \langle -\hbn' | { \bm \sigma} |\hbn'\rangle  = 1 + \hbn \cdot \hbn'$.
Crucially, the overlap between coherent states also simplifies to $\langle \hbn | \hbn' \rangle =  \sqrt{ (1 + \hbn \cdot \hbn')/2}$. The complex phase factor that is found in calculating this overlap from the coherent states, as defined in Eq.~(\ref{eq:CoherentStates}), is absent, having been absorbed into $\xi_{2i}$. 

It is this phase factor (with a phase angle proportional to the solid angle between $\hbn$, $\hbn'$ and $\hat {\bm z}$) that ultimately generates the coherent state contribution to the Berry phase. It is transferred, therefore, to a Berry phase arising from the dynamical fixing of the rotation $\xi_{2i}\equiv\xi_{2i}(\hbn_{2i-1}(t)\times \hbn_{2i}(t))$ (see the first term in Eq.~(\ref{eq:BerryPhaseSup}) below). In particular, we note that fixing $\xi_{2i}$ to its optimal value in this way does not alter the topological contribution to the Berry phase.

\section{Derivation of the non-linear sigma model}
\label{app:SigmaModel}

The central result of this letter is to show how the non-linear sigma model action and the Wess-Zumino topological term describing the long-wavelength physics of the $J_1$-$J_2$ chain can be connected to the spin wavefunction via the MPS ansatz.
Although we have given all of the essential steps in the main text, we include a more detailed derivation here. As mentioned in the main text, in order to map the dynamics of the MPS ansatz to the field theory, we use the restricted ansatz where consecutive bonds are not entangled.
We have
\eqn{
\mS = \int_0^{\beta} d\tau \bra{\Psi}\ket{\partial_{\tau}\Psi} + \mH,
}
where
\eqa{
\mH = J_1 \sum_{2i} \bigl(-w^x_{2i} + (1 + w_{2i}^x)\hbn_{2i-1}\cdot\hbn_{2i} + w^z_{2i}w^z_{2i+2}\hbn_{2i}\cdot\hbn_{2i+1} \bigr)
+ J_2 \sum_{2i} w^z_{2i}w^z_{2i+2}(\hbn_{2i-1}\cdot\hbn_{2i+1} + \hbn_{2i}\cdot\hbn_{2i+2}),
}
and
\eqa{
\mS_B = \int d\tau \bra{\Psi}\ket{\partial_\tau\Psi} = iS\int d\tau \biggl[\sum_{i}\cos\chi_{i}\,\bA(\hbn_{i})\cdot\dot{\hbn}_{i}
+ \sum_{2i}\dot \xi_{2i}\left( 1-\cos\chi_{2i} \right) \biggr]
\label{eq:BerryPhaseSup}
}
where $\bA(\hbn) = \frac{-\cos\theta}{\sin\theta} \hbf$ is a vector potential which generates the spin-coherent Berry phase. Note that we have simply set $\psi_{2i+1} = 0$ in the Hamiltonian and in the $\xi$-term of the Berry phase -- the same continuum limits are recovered by instead setting $\psi_{2i} = 0$. The topological term, however, will require more care, and we will deal with that separately in \S\ref{app:TopTerms}.

\subsection{Haldane's mapping and continuum limit}
To derive the continuum limit of this action, we adapt Haldane's mapping \cite{auerbach2012interacting}, and write
\eqn{
\hbn_i = (-1)^i \hbm_i \sqrt{1 - \bl_i^2} + \bl_i, \;\;\; \hbm_i^2 = 1, \;\;\; \hbm_i\cdot\bl_i = 0, \;\;\; \bl_i^2 \ll 1.
\label{eq:HaldaneMappingSup}
}
where $\hbm_i$ is the slowly-varying N\'eel field, and $\bl_i$ is the transverse canting field which captures any fast fluctuations of $\hbn_i$. 
Now, to second-order in $\bl$, we have
\eqa{
\hbn_i\cdot\hbn_j \sim (-1)^{i + j}\hbm_i\cdot\hbm_j + \bl_i\cdot\bl_j - \frac{1}{2}(-1)^{i + j}(\bl_i^2 + \bl_j^2)
+ (-1)^j \bl_i\cdot\hbm_j + (-1)^i\hbm_i\cdot\bl_j + \mathcal{O}(\bl^3). 
}
We will, shortly, approximate the differences of the slow $\hbm$-field by derivatives. We cannot take the continuum limit of the fast $\bl$-field, though the cross-terms turn out to be negligible in the Hamiltonian \cite{auerbach2012interacting}. At this order, then (and dropping the cross terms), the Hamiltonian becomes
\eqa{
\mH &= \sum_{2i} -J_1w^x_{2i} -J_1(1 + w^x_{2i})\hbm_{2i-1}\cdot\hbm_{2i} - J_1 w_{2i}^z w_{2i+2}^z \hbm_{2i}\cdot\hbm_{2i+1} + J_2 w_{2i}^z w_{2i+2}^z (\hbm_{2i-1}\cdot\hbm_{2i+1} + \hbm_{2i}\cdot\hbm_{2i+2}) \nn \\
&+ \sum_{2i} \biggl[ J_1(1 + w_{2i}^x)\left(\frac{\bl_{2i-1}^2}{2} + \bl_{2i-1}\cdot\bl_{2i} + \frac{\bl_{2i}^2}{2}\right) + J_1 w_{2i}^z w_{2i+2}^z \left(\frac{\bl_{2i}^2}{2} + \bl_{2i}\cdot\bl_{2i+1} + \frac{\bl_{2i+1}^2}{2}\right) \nn \\
&\;\;\;\;\;\;\;\;\; + J_2w_{2i}^z w_{2i+2}^z\left( - \frac{\bl_{2i-1}^2}{2} + \bl_{2i-1}\cdot\bl_{2i+1} - \frac{\bl_{2i+1}^2}{2} - \frac{\bl_{2i}^2}{2} + \bl_{2i}\cdot\bl_{2i+2} - \frac{\bl_{2i+2}^2}{2}\right) \biggr].
\label{eq:HamiltonianSecondOrder}
}
The first line in the above contains only the slow fields, and so we can proceed to the continuum limit. There is a subtlety, however, regarding the appropriate continuum limit for the N\'eel field---there is a much higher stiffness against gradients of $\hbm$ on the entangled bonds. Essentially, this causes the magnetisation to fluctuate on a scale of two lattice spacings---the same scale as for the entanglement. 

More explicitly, let us, temporarily, transform to sum and differences of the N\'eel fields across the entangled bonds,
\eqn{
\m_{2i}^{(+)} = \frac{\hbm_{2i-1} + \hbm_{2i}}{2}, \;\;\; \m_{2i}^{(-)} = \hbm_{2i-1} - \hbm_{2i}.
}
But, since the entangled bonds are stiff, we have, approximately,
\eqn{
\hbm_{2i-1} \approx \m_{2i}^{(+)}, \;\;\; \hbm_{2i} \approx \m_{2i}^{(+)}, \;\;\; \m_{2i}^{(-)} \approx 0, \;\;\; |\m_{2i}^{(+)}|^2 \approx 1.
} 
This implies the following continuum limits:
\eqa{
w_{2i}^z w_{2i+2}^z &\sim (w^z)^2 + 2w^z \pd_x^2 w^z \mapsto (w^z)^2 - 2(\pd_x w^z)^2 \nn \\
\hbm_{2i-1}\cdot\hbm_{2i} &\sim \m^{(+)}_{2i}\cdot\m^{(+)}_{2i} \sim 1, \nn \\ 
\hbm_{2i}\cdot\hbm_{2i + 1} &\sim \m^{(+)}_{2i}\cdot\m^{(+)}_{2i + 2} \sim 1 - 2(\pd_x \m^{(+)})^2 \sim 1 - 2(\pd_x \hbm)^2 \nn \\
\hbm_{2i - 1}\cdot\hbm_{2i + 1} &\sim \hbm_{2i}\cdot\hbm_{2i + 2} \sim \m^{(+)}_{2i}\cdot\m^{(+)}_{2i + 2} \sim 1 - 2(\pd_x \m^{(+)})^2 \sim 1 - 2(\pd_x \hbm)^2,
}
where we convert from $\pd_x^2 (\cdot) \mapsto -(\pd_x \cdot)^2$ using integration by parts, and the lattice spacing has been set to unity. Then the part of Eq.~\eqref{eq:HamiltonianSecondOrder} that contains only the slow fields becomes
\eqa{
...\;&\sim \int dx \biggl[-\frac{J_1}{2} - J_1 w^x - \left(\frac{J_1}{2} - J_2 \right) w_z^2 + (J_1 - 2J_2) \Bigl( (\pd_x w^z)^2 + w_z^2 (\pd_x \hbm)^2 \Bigr)\biggr].
}

\subsection{Berry phase and kinetic terms}
We now turn to the kinetic terms. Again, applying Eq.~\eqref{eq:HaldaneMappingSup}, we have
\eqa{
\mS_B = \int d\tau \bra{\Psi}\ket{\partial_\tau\Psi} = iS\int d\tau \biggl[\sum_{i}(-1)^i \dot{\phi}_i\cos\theta_i\cos\chi_i + \cos\chi_{i}\,\left(\hbm_{i}\times\pd_\tau\hbm_{i}\right)\cdot\bl_{i}
+ \sum_{2i}\dot \xi_{2i}\left( 1-\cos\chi_{2i} \right) \biggr]
\label{eq:BerryPhaseSup}
}
where $\theta$ and $\phi$ are the standard spherical co-ordinates for $\hbm$. The first term in the above is the topological term, which we will discuss shortly in \S\ref{app:TopTerms}. The remaining terms involve the fast fields $\bl$ and $\xi$.

Integrating out these fast fields yields the kinetic terms. We start with $\xi$, for which the corresponding part of the full action (Berry phase and Hamiltonian) is, in the continuum limit,
\eqa{
\mS_\xi &= \int d\tau dx \left[\frac{i}{4}(1 - \cos\chi)(\pd_\tau\xi) - J_1 \sin\chi \cos\xi \right] \nn \\
&\sim \int d\tau dx \left[+\frac{i}{4}(\pd_\tau\cos\chi)\xi - J_1|\sin\chi| + \frac{J_1}{2}|\sin\chi|\xi^2 \right] \nn \\
&= \int d\tau dx \left[ \frac{J_1}{2}|\sin\chi|\left( \xi + \frac{i}{4 J_1 |\sin\chi|} (\pd_\tau\cos\chi)\right)^2 + \frac{1}{32J_1|\sin\chi|}(\pd_\tau\cos\chi)^2 - J_1|\sin\chi| \right],
}
which, after completing the Gaussian integrals over the transformed $\xi$-field, and approximating ${|\sin\chi| \approx 1}$ in the coefficient of the entanglement field, contributes
\eqn{
\int d\tau dx \biggl[-J_1|\sin\chi| + \frac{1}{32J_1}(\pd_\tau\cos\chi)^2 \biggr]
}
to the total action.

We now turn to the fast fluctuations of the magnetisation, $\bl$. Since we are expanding around the singlet covers, we neglect the terms $\sim\cos^2\chi \, \bl^2$ in Eq.~\eqref{eq:HamiltonianSecondOrder}. We thus obtain:
\eqa{
\mS_{\bl} &= \int d\tau \sum_q \frac{i}{2}\cos\chi \left(\hbm \times \pd_\tau\hbm \right)_q \cdot \bl_{-q} + J_1 \omega_q f(\chi) \bl_q\cdot\bl_{-q} \nn \\
&= \int d\tau \sum_q \biggl[ J_1\omega_q f(\chi) \left(\bl_q + \frac{i}{4J_1\omega_q f(\chi)}\cos\chi \left(\hbm \times \pd_\tau\hbm \right)_q \right)\cdot\left(\bl_{-q} + \frac{i}{4J_1\omega_q f(\chi)}\cos\chi \left(\hbm \times \pd_\tau\hbm \right)_{-q} \right) \nn \\
&\;\;\;\;\;\;\;\;\;\;\;\;\;\;\;\;\;\;+ \frac{1}{16J_1\omega_q f(\chi)} \cos^2\chi \left(\hbm \times \pd_\tau\hbm \right)_q\cdot\left(\hbm \times \pd_\tau\hbm \right)_{-q} \biggr],
}
where $f(\chi) = 1 + |\sin\chi|$ and $\omega_q = \frac{1}{2}(1 + \cos q)$, and we have assumed that $\chi$ varies much more slowly than $\bl$, such that its momentum-dependence can be neglected when taking the Fourier transform of $\bl$. Then, completing the integrals of the transformed $\bl$-field, approximating ${|\sin\chi| \approx 1}$ in the coefficient, and replacing $\omega_q$ by its zero momentum value (again, justified because the variation of $\bl$ is much faster than that of $\hbm$), we have
\eqn{
\int d\tau dx \; \frac{1}{32J_1}\cos^2\chi (\pd_\tau\hbm)^2,
}
where we have made use of the identity $(\hbm \times \pd \hbm)^2 = (\pd \hbm)^2$. 

Finally, we switch to the standard angular co-ordinates on $S^3$. As in the main text, we define $\alpha = -\chi + \pi/2$, in terms of which $(w^z)^2 = \sin^2\alpha$ and $(\pd w^z)^2 \sim (\pd\alpha)^2$. The full action, then, becomes
\eqn{
\mS = i\Omega + \int d\tau dx \biggl[ \frac{1}{32J_1}\biggl((\pd_\tau\alpha)^2 + \sin^2\alpha (\pd_\tau \hbm)^2\biggr) + (J_1 - 2J_2)\biggl( (\pd_x \alpha)^2 + \sin^2\alpha(\pd_x\hbm)^2 \biggr) + V(\alpha) \biggr],
}
with the dimerisation potential
\eqn{
V(\alpha) = -\frac{J_1}{2} - J_1|\cos\alpha| - \left(\frac{J_1}{2} - J_2\right)\sin^2\alpha \approx -\frac{3J_1}{2} + \frac{J_2}{2} - \frac{J_2}{2}\cos2\alpha
}
(the approximation preserves $V(\alpha)$ to second order around the singlet covers $\alpha = 0$ or $\pi$). If $V(\alpha)$ is relevant, we are in the dimer phase; if $V(\alpha)$ is irrelevant we are in the critical phase, with the $SO(4)$ symmetry of the joint-order parameter emergent in the infrared, and the topological term protecting the gaplessness.

\section{The topological term \label{app:TopTerms}}

Finally, let us show how the Wess-Zumino term is encoded in the Berry phase of the MPS ansatz and emerges directly in the continuum limit. 
After applying the Haldane mapping \eqref{eq:HaldaneMappingSup}, the Berry phase \eqref{eq:BerryPhaseSup} contains a term
\eqn{
i\Omega = i S \int d\tau \sum_i (-1)^i\dot{\phi}_i\cos\theta_i\cos2\psi_i\cos2\psi_{i+1},
\label{eq:topological_term_lattice}
}
where the factor $(-1)^i$ appears because $\theta$ and $\phi$ are the co-ordinates of the N\'eel field $\hbm$, and we have not yet restricted the entanglement.

Now, recall that in the restricted ansatz (from which the microscopic order parameter (8) is constructed), consecutive bonds are not entangled -- that is, if $\psi_i \neq 0$, then $\psi_{i-1} = \psi_{i+1} = 0$. As discussed in the main text, we enforce this by introducing an angle $\chi_i$, defined by
\eqn{
\begin{cases}
    2\psi_{2i+1} = \chi_{2i} \Theta(\chi_{2i}) = \chi_{2i+1} \Theta(\chi_{2i+1}) \vspace{0.1cm}\\
    2\psi_{2i}= -\chi_{2i}\Theta(-\chi_{2i}) = -\chi_{2i-1}\Theta(-\chi_{2i-1}).
\end{cases}
}
That is, $\chi_{2i} > 0$ means that the bond $(2i, 2i+1)$ is entangled; $\chi_{2i} < 0$ means that the bond $(2i-1, 2i)$ is entangled. 
Most of the $\chi_i$ are not independent:
for example, if $\chi_{2i} > 0$, then $\chi_{2i+1} = \chi_{2i}$; and if $\chi_{2i} < 0$, then $\chi_{2i-1} = \chi_{2i}$. We show an illustrative example of a configuration of $\chi$ in Fig.~\ref{fig:chi_example}.


To take the continuum limit we have to rewrite Eq.~\eqref{eq:topological_term_lattice} in terms of $\chi$. 
Consider first a region where $\chi_{2i} > 0$. Then the bonds $(2i, 2i+1)$ are entangled, $\chi_{2i} = \chi_{2i+1} = 2\psi_{2i+1}$, and the Berry phase is
\eqn{
i\Omega_{\chi>0} =
i S \int d\tau \sum_{2i} \cos\chi_{2i} \left( \dot\phi_{2i} \cos\theta_{2i} -  \dot\phi_{2i+1} \cos\theta_{2i+1} \right).
}
This has the form of a Haldane-like term modulated by the entanglement field; but this is incomplete in the presence of domain walls, where $\chi$ changes sign. In a region where $\chi_{2i} < 0$, we instead have
\eqn{
i\Omega_{\chi<0} =
i S \int d\tau \sum_{2i} \cos\chi_{2i} \left( \dot\phi_{2i} \cos\theta_{2i} -  \dot\phi_{2i-1} \cos\theta_{2i-1} \right);
}
the shift in the sites to which the spin-coherent terms refer is important -- it changes the sign of their derivative in the continuum limit. Finally, if the sign of $\chi$ changes -- that is, if there is a domain wall -- there will be a lattice site where $\chi = 0$, and the spin is not entangled with either neighbour.

\begin{figure}
\begin{tikzpicture}
    \foreach \x in {0,1,2,3,4,5,6,7,8,9,10,11,12,13} {
        \fill[black] (\x,0) circle (2pt);
    }
    \draw[blue] (0,0) -- (1,0);
    \draw[blue] (2,0) -- (3,0);
    \draw[black, ->] (3.9, -0.3) -- (4.1, 0.3);
    \draw[red] (5,0) -- (6,0);
    \draw[red] (7,0) -- (8,0);
    \draw[black, <-] (8.9, -0.3) -- (9.1, 0.3);
    \draw[blue] (10,0) -- (11,0);
    \draw[blue] (12,0) -- (13,0);

    \draw[black, dotted] (-0.5,0) -- (-0.2,0);
    \draw[black, dotted] (13.2,0) -- (13.5,0);

    \draw (0.5,-0.2) node[below] {$\chi_e = \chi_o$};
    \draw (2.5,-0.2) node[below] {$\chi_e = \chi_o$};
    \draw (1.5, 0.3) node[above] {$\chi > 0$};
    \draw (4,-0.2) node[below] {$\chi_e = 0$};
    \draw (5.5,-0.2) node[below] {$\chi_o = \chi_e$};
    \draw (7.5,-0.2) node[below] {$\chi_o = \chi_e$};
    \draw (6.5, 0.3) node[above] {$\chi < 0$};
    \draw (9,-0.2) node[below] {$\chi_o = 0$};
    \draw (10.5,-0.2) node[below] {$\chi_e = \chi_o$};
    \draw (11.5, 0.3) node[above] {$\chi > 0$};
    \draw (12.5,-0.2) node[below] {$\chi_e = \chi_o$};
\end{tikzpicture}
\caption{An example configuration of the entanglement field $\chi$ (the leftmost site is even), showing how the type of singlet order (even-odd or odd-even) is encoded in the sign of $\chi$, and that domain walls in the singlet order correspond to free spins.}
\label{fig:chi_example}
\end{figure}

Putting all this together, then, we can write the Berry phase in terms of $\chi$ as
\eqa{
&i\Omega = i\Omega^{(\mathrm{H})} + i\Omega^{(\mathrm{DW})}; \nn \\[0.5cm]
&i\Omega^{(\mathrm{H})} = iS\int d\tau \sum_{2i} \biggl[ \Theta(\chi_{2i}) \cos\chi_{2i}\left(-\dot{\phi}_{2i+1}\cos\theta_{2i+1} +\dot{\phi}_{2i}\cos\theta_{2i} \right) \nn \\
&\;\;\;\;\;\;\;\;\;\;\;\;\;\;\;\;\;\;\;\;\;\;\;\;\;\;\;\;\;\;\;\;\;\;+ \Theta(-\chi_{2i}) \cos\chi_{2i}\left(\dot{\phi}_{2i}\cos\theta_{2i} - \dot{\phi}_{2i-1}\cos\theta_{2i-1} \right)\biggr], \nn \\[0.2cm]
&i\Omega^{(\mathrm{DW})} = iS\int d\tau \sum_{2i} \biggl[ \delta_{\chi_{2i}, 0}\,\dot{\phi}_{2i}\cos\theta_{2i} - \delta_{\chi_{2i+1},0}\,\dot{\phi}_{2i+1}\cos\theta_{2i+1} \biggr],
}
where we have split $i\Omega$ into `Haldane-like' and `domain wall-like/free spin' terms.

The continuum limit of $i\Omega^{(\mathrm{H})}$ is relatively straightforward -- though we should be careful to note that when $\chi$ changes sign the spin coherent terms shift by one site, changing the sign of their derivative. We thus obtain:
\eqa{
i\Omega^{(\mathrm{H})} &\sim \frac{iS}{2} \int d\tau dx (-\Theta(\chi) + \Theta(-\chi)) \cos\chi \left(\pd_\tau\phi\;\pd_x\cos\theta - \pd_x\phi\;\pd_\tau\cos\theta \right) \nn \\
&= -\frac{iS}{2} \int d\tau dx \; \mathrm{sgn}\chi \cos\chi \left(\pd_\tau\phi\;\pd_x\cos\theta - \pd_x\phi\;\pd_\tau\cos\theta \right).
}

The continuum limit of the domain wall-type terms is more subtle. It will be easier to see if we first consider the case where the spins are in a perfectly antiferromagnetic configuration, i.e., $\dot{\phi}_i\cos\theta_i = \dot{\phi}\cos\theta$, so we can take that outside the sum. In that case we have
\eqn{
i\Omega^{(\mathrm{DW})} = iS\int d\tau \;\dot{\phi}\cos\theta \sum_{2i} \biggl[ \delta_{\chi_{2i}, 0} - \delta_{\chi_{2i+1},0}\biggr].
}
We observe that the sum simply counts the number of unentangled even sites, minus the number of unentangled odd sites. Equivalently, cf. Fig.~\ref{fig:chi_example}, it counts the number of times $\chi$ changes from positive to negative, minus the number of times $\chi$ changes from negative to positive. This latter observation tells us how to take the continuum limit:
\eqn{
\sum_{2i} \biggl[ \delta_{\chi_{2i}, 0} - \delta_{\chi_{2i+1},0}\biggr] = -\int dx \;\pd_x \Theta(\chi(x)).
}
Moving the spin coherent terms back inside the integral, then, we have
\eqa{
i\Omega^{(\mathrm{DW})} &\sim iS\int d\tau dx\; \left(-\pd_x \Theta(\chi)\right)\cos\theta\;\pd_\tau\phi \nn \\
&= iS\int d\tau dx\; \Theta(\chi)\left(\pd_\tau\phi\;\pd_x\cos\theta - \pd_x\phi\;\pd_\tau\cos\theta \right),
}
where we have integrated by parts to obtain the second line. Putting it all together, then,
\eqn{
i\Omega \sim iS \int d\tau dx \; \left(\Theta(\chi) -\frac{\mathrm{sgn}\chi\cos\chi}{2}\right) \left(\pd_\tau\phi\;\pd_x\cos\theta - \pd_x\phi\;\pd_\tau\cos\theta \right).
}
This is equivalent to the Wess-Zumino term, in the sense that both terms produce the same functional weight in the path integral.
To compare directly to the Wess-Zumino term, it is useful to define
\eqn{
f(\chi) = \Theta(\chi) - \frac{\mathrm{sgn}\chi\cos\chi}{2}.
\label{eq:f_chi_sup}
}

To prove the equivalence, consider the standard formulation of the Wess-Zumino term,
\eqn{
\Omega_{\mathrm{WZ}} = \frac{4\pi S k}{2\pi^2}
\int_0^1 d\zeta \int d\tau dx
\,\epsilon_{abcd} 
u^a \partial_\zeta u^b \partial_\tau u^c \partial_x u^d,
}
where the factor of $2 \pi^2$ in the denominator is the area of $S^3$, and $u(\zeta, \tau, x)$ is an arbitrary extension of the field $u(\tau,x) = (\sin\chi, \cos\chi\,\hbm) \equiv (u^0, \bm u)$ satisfying 
\begin{eqnarray}
u(\zeta=0,\tau,x) &=& (1,0,0,0) \nonumber\\
u(\zeta=1,\tau,x) &=& u(\tau,x).
\end{eqnarray}
The purpose of the extension is simply to allow the term to be written in a co-ordinate free way. 
Without loss of generality, we can assume that only $\chi$ depends on the extension co-ordinate, and the integral over $\zeta$ can then be explicitly performed. We thus have 
\eqn{
\Omega_{\mathrm{WZ}} = k S \int d\tau dx \left(\frac{2\chi + \sin2\chi}{2\pi} - \frac{1}{2}\right) \left(\pd_\tau\phi\;\pd_x\cos\theta - \pd_x\phi\;\pd_\tau\cos\theta \right),
}
i.e., comparing to the definition of $f(\chi)$ \eqref{eq:f_chi_sup},
\eqn{
f_\mathrm{WZ}(\chi) = \frac{2\chi + \sin2\chi}{2\pi} - \frac{1}{2},
}
where we have now set the level to $k = 1$. At first glance, since $f(\chi) \neq f_{\mathrm{WZ}}(\chi)$, it would appear that the two terms are not the same. However, this is a topological term -- any homeomorphic deformation of the target space results in the same value; explicitly, then, these integrals are equal if $f$ and $f_{\mathrm{WZ}}$ are both monotonic and equal to each other at $\chi = -\pi/2$ and $\chi = \pi/2$. We note also that we may shift $f$ by any integer value, since such a shift corresponds to a term valued in $4\pi i S\mathbb{Z}$,
and so does not change the 
weight of any configuration in the path integral. We observe that $f$
and $f_{\mathrm{WZ}}$ satisfy these conditions -- and so the continuum limit of the Berry phase of the MPS ansatz is equivalent to the Wess-Zumino term.

\end{document}